\documentclass[aps,pra,reprint,showpacs,groupedaddress]{revtex4-1}
\usepackage{amsmath,amssymb,mathrsfs,amsfonts}
\usepackage{epsfig}
\usepackage{bm}
\usepackage{color}
\usepackage[dvipdfm,
            pdfstartview=FitH,
            CJKbookmarks=true,
            bookmarksnumbered=true,
            bookmarksopen=true,
            linktocpage=true,
            colorlinks=true, 
            pdfborder=001,   
            citecolor=blue,  
            urlcolor=blue,   
            linkcolor=blue,  
            anchorcolor=blue,
            ]{hyperref}      

\begin{document}
  \title{Non-Markovian qubit dynamics in nonequilibrium environments}
  \author{Xiangji Cai}
   \email{xiangjicai@foxmail.com}
  \affiliation{School of Science, Shandong Jianzhu University, Jinan 250101, China}

\begin{abstract}
  We theoretically study the non-Markovian dynamics of qubit systems coupled
  to nonequilibrium environments with nonstationary and non-Markovian statistical properties.
  The reduced density matrix of the single qubit system satisfies a closed third-order
  differential equation with all the higher-order environmental correlations taken into account
  and the reduced density matrix of the two qubit system can be expressed as the Kraus representation
  in terms of the tensor products of the single qubit Kraus operators.
  We derive the relation between the entanglement and nonlocality of the two qubit system which are
  both closely associated with the decoherence function.
  We identify the threshold values of the decoherence function to ensure the existences of the concurrence and nonlocal quantum correlations for a given evolution time when the two qubit system is initially prepared in the composite Bell states and the extended Werner states, respectively.
  It is shown that the environmental nonstationary feature can suppress the decoherence and disentanglement
  dynamics and can reduce the coherence and entanglement revivals in non-Markovian dynamics regime.
  In addition, it is shown that the environmental non-Markovian feature can enhance the coherence revivals in the single decoherence dynamics and the entanglement revivals in the two qubit disentanglement dynamics, respectively.
  Furthermore, the environmental nonstationary and non-Markovian features can enhance the nonlocality of the two qubit system.
\end{abstract}

\maketitle

\section{Introduction}
\label{sec:intr}

Coherence and entanglement are two basic quantum features of nonclassical systems which play vital roles in quantum mechanical community as specific resources ranging from fundamental questions to wide applications in quantum computing, quantum metrology and quantum information science~\cite{RevModPhys.81.865,PhysRevLett.117.020402,RevModPhys.89.041003,PhysRep.762.1,RevModPhys.91.025001}.
It is known that any quantum system loses quantum features during time evolution resulting from the unavoidable couplings between the system and the environments.
The study of decoherence and disentanglement dynamics of open quantum systems can help us further expand the understanding of the environmental effects on the dynamical evolution of the quantum systems and the real origins of loss of quantum features and quantum-classical transition, which has potential applications in preserving quantum features against the environmental noise and in realizing quantum manipulation and control and quantum measurement~\cite{Breuerbook,Schlosshauerbook,PhysRep.831.1,PhysRep.838.1,PhysRevLett.91.066801,PhysRevLett.94.066803,PhysRevB96.235417,PhysRevB101.174302}.

In recent decade, it has attracted increasing attention to study theoretically the dynamics of open quantum systems beyond the framework of Markov approximation~\cite{PhysRevLett.82.1801,PhysRevLett.100.180402,PhysRevLett.103.210401,PhysRevLett.105.050403,PhysRevLett.109.170402,
PhysRevLett.112.120404,RepProgPhys.77.094001,PhysRevLett.112.210402,RevModPhys.88.021002,RevModPhys.89.015001},
and there have been well established theoretical approaches to study the non-Markovian dynamics of open quantum systems which usually assume that the environments are in equilibrium and exhibit both Markovian and stationary statistical properties within the framework of classical and quantum treatments~\cite{JPhysSocJpn.9.316,JPhysSocJpn.9.935,
PhysRevLett.88.228304,PhysRevLett.94.167002,PhysRevLett.99.160502,PhysRevA77.032342,PhysRevA78.060302,Entropy17.2341,
PhysRevLett.118.140403,PhysRevA95.052126,PhysRevA97.042126,RevModPhys.86.361,JChemPhys.142.094106,JChemPhys.148.014103}.
However, it is quite difficult to obtain the exact dynamical evolution equation for open quantum systems and only a few physical models can be exactly solved due to the indispensable influences of the higher-order environmental correlations.
Thus, it is extremely meaningful to derive the exact quantum dynamical equation with all the higher-order environmental correlations effects taken into account, which can help us better understand the real decoherence process induced by the environments and has potential applications in quantum coherent manipulation and control.

Recently, it has been observed experimentally the nonequilibrium feature of the environments in many crucial dynamical processes~\cite{PhysRevLett.100.176805,NatPhys.8.522}.
In these processes, the environmental nonequilibrium states caused by the interaction with the quantum systems can not reach stationary in time~\cite{JChemPhys.133.241101,JPhysB45.154008,JChemPhys.139.024109,PhysRevA87.032338,PhysRevA91.042111} which corresponds to that the environments around the quantum systems are out of equilibrium and
exhibit nonstationary and non-Markovian statistical properties.
It has increasingly drawn much attention to study the environmental nonequilibrium effects on
quantum coherence due to the significant role in the dynamical evolution of the open quantum systems,
and the theoretical results show that nonequilibrium environments cause the energy levels shift of the quantum system and delay the transition critical time of decoherence from classical to quantum~\cite{PhysRevA94.042110,PhysRevA95.052104,JChemPhys.149.094107,EurophysLett.125.30007,NewJPhys.22.033039}.
Meanwhile, some other important physical questions arise naturally and should be addressed: How do the nonstationary and non-Markovian statistical properties of the nonequilibrium environments influence the disentanglement dynamics of open quantum systems, respectively?
Can we find a close relation between the local decoherence and nonlocal entanglement of open quantum systems in nonequilibrium environments?
And what is the influence of the environmental nonequilibrium feature on the quantum nonlocality?

In this paper, we theoretically study the non-Markovian dynamics of qubit systems interacting with nonequilibrium environments which display nonstationary and non-Markovian random telegraph noise (RTN) properties.
It is shown that the nonstationary statistical property of the environments give rise to nonzero environmental odd-order correlation functions which play an essential role in the unitary dynamical evolution of the qubit system.
We derive the closed third-order differential equation for the single qubit reduced density matrix with all the higher-order environmental correlations taken into account.
By means of Laplace transformation, we derive analytically the exact expression of the single qubit reduced density matrix.
The two qubit system consists of two noncoupling identical single qubits independently interacting with its local nonequilibrium environment, of which the reduced density matrix can be expressed as the Kraus representation in terms of the tensor products of the single qubit Kraus operators.
We derive the relation between entanglement quantified by the concurrence and nonlocality characterized by the Bell function closely associated with the decoherence function.
We identify the threshold values of the decoherence function to ensure the existences of the concurrence and nonlocal quantum correlations at an arbitrary evolution time for the two qubit system prepared initially in the composite Bell states and the extended Werner states, respectively.
It is shown that the environmental nonstationary feature can suppress both the decoherence and disentanglement dynamics and that it can reduce the coherence and entanglement revivals in non-Markovian dynamics regime.
In addition, the environmental nonstationary feature can enhance the nonlocality of the two qubit system.
It is shown that the environmental non-Markovian feature can increase the coherence and entanglement revivals in the single and two qubit dynamics, respectively.
Furthermore, the environmental non-Markovian feature can also enhance the nonlocality of the two qubit system.

This paper is organized as follows. In Sec.~\ref{sec:theo} we introduce the theoretical framework of
non-Markovian qubit dynamics in nonequilibrium environments.
We employ the non-Markovianity and decoherence function to quantify the non-Markovian single qubit decoherence dynamics and we employ the non-Markovianity, concurrence and Bell function to describe the non-Markovian two qubit disentanglement dynamics in nonequilibrium environments.
In Sec.~\ref{sec:result}, we discuss the numerical results of the non-Markovian decoherence dynamics of a single qubit system and disentanglement dynamics of a two noninteracting qubit system coupled to nonequilibrium environments with nonstationary and non-Markovian statistical properties, respectively.
In Sec.~\ref{sec:con} we give the conclusions from the present study.

\section{Qubit dynamics in nonequilibrium environments}
\label{sec:theo}
\subsection{Decoherence dynamics of a single qubit system}
We consider a single qubit system $S$ with the states $|0\rangle$ and $|1\rangle$ interacting with a nonequilibrium environment $E$. The environmental initial state $\rho_{E}(0)$ is in nonequilibrium corresponding to the nonstationary statistical property of the environment.
In the interaction picture with respect to the single qubit system Hamiltonian $H_{S}=(\hbar/2)\omega_{S}\sigma_{z}$ and the environment Hamiltonian $H_{E}$, the Hamiltonian of the total system $S+E$ can be written as~\cite{PhysRevLett.116.150503}
\begin{equation}
\label{eq:sinintHam}
  H_{SE}(t)=\frac{\hbar}{2}\epsilon(t)\sigma_{z},
\end{equation}
where $\omega_{S}$ is the frequency difference of the qubit system, $\sigma_{z}$ denotes the Pauli matrix in the single qubit basis $\mathcal{B}_{S}=\{|0\rangle, |1\rangle\}$, and the environmental noise $\epsilon(t)$ is subject to a stochastic process in the presence of a classical environment whereas it is a time dependent operator in the presence of a quantum environment.

The qubit system undergoes pure decoherence during its dynamical evolution due to the fact that $[H_{S},H_{SE}(t)]=0$,
and the environmental decoherence effect on the qubit system is reflected in the dynamics of the coherence element.
The dynamical evolution of the density matrix of the total system is described by the Liouville equation $d\rho_{SE}(t)/dt=-(i/\hbar)[H_{SE}(t),\rho_{SE}(t)]$.
By taking a partial trace over the degrees of freedom of the environment,
the reduced density matrix of the single qubit system can be written as
\begin{equation}
\label{eq:sinreddenmat}
\begin{split}
  \rho_{S}(t)&=\left\langle U_{SE}\bm(t;\epsilon(t)\bm)\rho_{SE}(0)U_{S}^{\dag}\bm(t;\epsilon(t)\bm)\right\rangle\\
  &=\sum_{\mu=1}^{2}K_{S\mu}(t)\rho_{S}(0)K_{S\mu}^{\dag}(t),
\end{split}
\end{equation}
where $\langle\cdots\rangle=\mathrm{Tr}_{E}(\cdots\rho_{E})$ represents a trace over the environmental degrees of freedom
with the environment state $\rho_{E}(t)$, $U_{SE}\bm(t;\epsilon(t)\bm)=\exp\big[-(i/\hbar)\int_{0}^{t}dt'H_{SE}(t')\big]$
denotes the unitary time evolution operator in terms of the $H_{SE}(t)$ in Eq.~(\ref{eq:sinintHam}) and the single qubit Kraus operators $K_{S\mu}$ satisfy
\begin{equation}
\label{eq:sinqubKra}
  K_{S1}(t)=\left
   (\begin{array}{cc}
    1&0\\
    0&D(t)
   \end{array}
   \right),K_{S2}(t)=\left
   (\begin{array}{cc}
    0&0\\
    0&\sqrt{1-|D(t)|^{2}}
   \end{array}
   \right),
\end{equation}
with the decoherence function $D(t)=e^{-i\Phi(t)-\Gamma(t)}$ in terms of the time-dependent phase angle $\Phi(t)$ and decay factor $\Gamma(t)$ expressed respectively as
\begin{equation}
\label{eq:phadec}
\begin{split}
  \Phi(t)&=\sum_{n=1}^{\infty}\frac{(-1)^{n}}{(2n-1)!}\mathcal{C}_{2n-1}(t),\\
  \Gamma(t)&=\sum_{n=1}^{\infty}\frac{(-1)^{n}}{(2n)!}\mathcal{C}_{2n}(t),\\
  \mathcal{C}_{n}(t)&=\int_{0}^{t}dt_{1}\cdots\int_{0}^{t}dt_{n}C_{n}(t_{1},\cdots,t_{n}),
\end{split}
\end{equation}
based on the statistical characteristics of the environmental noise $\epsilon(t)$.
The $n$th-order cumulant $C_{n}(t_{1},\cdots,t_{n})$ depends on the environmental correlation function $\langle\epsilon(t_{1})\cdots\epsilon(t_{k})\rangle$ with $k\leq n$ and $\langle\cdots\rangle$ denoting an ensemble average with respect to the initial state of the environment $\rho_{E}(0)$.
The odd-order and even-order cumulants of the environmental noise characterize the degrees of the asymmetry and broadness of the distribution, respectively.
It is worth mentioning that the environmental noise $\epsilon(t)$ is stationary if and only if $[H_{E},\rho_{E}(0)]=0$ which corresponds to the environment in a certain equilibrium state initially, and in this case the qubit system undergoes no phase evolution since the odd-order cumulants of the environmental noise vanish~\cite{PhysRevLett.116.150503}.
For the general case $[H_{E},\rho_{E}(0)]\neq0$ corresponding to the environment in an initial nonequilibrium state, the statistical property of the environment is nonstationary and the environmental odd-order correlation functions play an essential role in the unitary dynamical evolution of the qubit system.

Based on the Kraus operators expression for the single qubit reduced density matrix in Eq.~(\ref{eq:sinreddenmat}),
the diagonal elements are time independent and the off diagonal elements are time dependent
\begin{equation}
\label{eq:elesinden}
\begin{split}
 \rho_{00}(t)&=\rho_{00}(0),\\
 \rho_{11}(t)&=1-\rho_{00}(t),\\
 \rho_{01}(t)&=\rho_{10}^{*}(t)=\rho_{01}(0)D(t).
\end{split}
\end{equation}
The dynamics of the single qubit system satisfies
a time-local quantum master equation as
\begin{equation}
\label{eq:masequ}
  \frac{d}{dt}\rho_{S}(t)=-\frac{i}{2}\phi(t)[\sigma_{z},\rho_{S}(t)]-\frac{1}{4}\gamma(t)[\sigma_{z},[\sigma_{z},\rho_{S}(t)]],
\end{equation}
where $\phi(t)=d\Phi(t)/dt$ represents the frequency shift caused by the odd-order environmental correlations and $\gamma(t)=d\Gamma(t)/dt$ denotes the decoherence rate induced by the even-order environmental correlations.
Due to the environmental decoherence, the diagonal elements of the qubit reduced density matrix do not evolve with time
while the off-diagonal elements decay monotonously (Markovian behavior) or non-monotonously (non-Markovian behavior).
The non-Markovianity quantifying the exchange of a flow of information between the single qubit system and environment can be, by taking the optimization over all pairs of initial states, expressed as~\cite{PhysRevLett.103.210401,PhysRevA81.062115}
\begin{equation}
\label{eq:sinnonMar}
\begin{split}
 \mathcal{N}_{S}&=\max\limits_{\rho_{S}^{1,2}(0)}\int_{\frac{d\mathcal{D}}{dt}>0}
 \frac{d}{dt}\mathcal{D}(\rho_{S}^{1}(t),\rho_{S}^{2}(t))dt\\
 &=-\int_{\gamma(t)<0}\gamma(t)|D(t)|dt,
\end{split}
\end{equation}
where $\mathcal{D}(\rho^{1},\rho^{2})=\frac{1}{2}\mathrm{tr}|\rho^{1}-\rho^{2}|$ denotes the trace distance between the states $\rho^{1}$ and $\rho^{2}$ and the optimal pair of initial states is chosen as the maximally coherent states $|\psi_{S}^{\pm}(0)\rangle=(|0\rangle\pm|1\rangle)/\sqrt{2}$.
The single qubit dynamics displays non-Markovian behavior if the decoherence rate $\gamma(t)$ takes negative values in some time intervals.

The environmental influences on open quantum systems have been well modeled by RTN in a wide variety of dynamical processes in physics, chemistry and biology,
such as, fluorescence process of single molecules~\cite{PhysRevLett.90.238305,PhysRevLett.90.120601}, decoherence and disentanglement processes in the presence of $1/f^{\alpha}$ noise~\cite{PhysRevB79.125317,PhysRevB90.054304,PhysRevB94.235433} and optical trapping and frequency modulation processes in quantum optics~\cite{Nature330.769,RepProgPhys.80.056002}.
Unlike the Gaussian noise of which all cumulants $C_{n}(t_{1},\cdots,t_{n})$ beyond $n=2$ are zero, the generalized RTN displays distinct non-Gaussianity. Its cumulants higher than second-order do not vanish and contribute to the dynamical decoherence of the qubit system.
We here consider that the environmental noise $\epsilon(t)$ exhibits with generalized RTN property which is non-Markovian and nonstationary.
The noise process transits randomly with an average transition rate $\eta$ between values $\pm1$ with the amplitude $\chi$ which describes the environmental coupling.
Its non-Markovian and nonstationary features are respectively characterized by a generalized master equation for the conditional probability with a memory kernel and by an initially nonstationary distribution.
For the generalized RTN process, the environmental non-Markovian feature is described by a generalized master equation for
the time evolution of the conditional probability~\cite{PhysRevE50.2668}
\begin{equation}
\label{eq:genmasequ}
  \frac{\partial}{\partial t}P(\varepsilon,t|\varepsilon',t')=\int_{t'}^{t} K(t-\tau)\eta TP(\varepsilon,\tau|\varepsilon',t')d\tau,
\end{equation}
where $K(t-\tau)$ is the memory kernel of the environmental noise and the conditional probability $P(\varepsilon,t|\varepsilon',t')$ and transition matrix $T$ are respectively expressed as
\begin{equation}
\label{eq:conprotramat}
   P(\varepsilon,t|\varepsilon',t')=
   \left
   (\begin{array}{c}
    P(+\chi,t|\varepsilon',t')\\
    P(-\chi,t|\varepsilon',t')
   \end{array}
   \right),
    T=
   \left
   (\begin{array}{cc}
    -1&1\\
    1&-1
   \end{array}
   \right).
\end{equation}
We consider the case that the environmental memory kernel is of an exponential form as $K(t-\tau)=\kappa e^{-\kappa(t-\tau)}$ with $\kappa$ denoting the memory decay rate.
The smaller the decay rate $\kappa$ is, the stronger the environmental non-Markovian feature is.
For the case $\kappa\rightarrow\infty$, namely, the memoryless case $K(t-\tau)=\delta(t-\tau)$, the environmental noise only exhibits Markovian feature.

By means of the Laplace transformation $\tilde{P}(\varepsilon,s|\varepsilon',t')=\int_{0}^{\infty}P(\varepsilon,t|\varepsilon',t')e^{-st}dt$, the conditional probability in Eq.~(\ref{eq:genmasequ}) can be analytically expressed as
\begin{equation}
\label{eq:conprotramat}
\begin{split}
   \tilde{P}(\varepsilon,s|\varepsilon',t')=&[sI_{2}-\tilde{K}(s)T]^{-1}P(\varepsilon,t'|\varepsilon',t')\\
   =&
   \frac{1}{2}\left
   (\begin{array}{cc}
    \frac{1}{s}+e^{-st'}\widetilde{\Omega}(s)&\frac{1}{s}-e^{-st'}\widetilde{\Omega}(s)\\
    \frac{1}{s}-e^{-st'}\widetilde{\Omega}(s)&\frac{1}{s}+e^{-st'}\widetilde{\Omega}(s)
   \end{array}
   \right)P(\varepsilon,t'|\varepsilon',t'),
\end{split}
\end{equation}
where $I_{2}$ is the $2\times2$ identity matrix, $\widetilde{K}(s)=\kappa/(s+\kappa)$ represents the Laplace transform of the environmental memory kernel, and $\widetilde{\Omega}(s)=1/[s+2\lambda\widetilde{K}(s)]$.
The environmental nonstationary feature arises from the initial distribution
\begin{equation}
\label{eq:inidis}
  P(\varepsilon,0)=\left
   (\begin{array}{c}
    P(+\chi,0)\\
    P(-\chi,0)
   \end{array}
   \right)=\frac{1}{2}\left
   (\begin{array}{c}
    1+a_{0}\\
    1-a_{0}
   \end{array}
   \right),
\end{equation}
where $a_{0}$ is the nonstationary parameter and $-1\leq a_{0}\leq1$. For the case $a_{0}=0$, the environment is in equilibrium and the environmental noise only displays stationary feature.

Based on the non-Markovian and nonstationary features, the statistical characteristics of the environmental noise $\epsilon(t)$ are described by the first and second-order moments
\begin{equation}
\label{eq:avecor}
\begin{split}
  M_{1}(t)&=\langle\epsilon(t)\rangle=a_{0}\chi\mathscr{L}^{-1}[\widetilde{\Omega}(s)],\\
  M_{2}(t,t')&=\langle\epsilon(t)\epsilon(t')\rangle=\chi^{2}\mathscr{L}^{-1}[e^{-st'}\widetilde{\Omega}(s)],
\end{split}
\end{equation}
where $\mathscr{L}^{-1}$ denotes the inverse Laplace transform. According to the Bayes' theorem, the environmental higher-order moments satisfy the factorization
\begin{equation}
\label{eq:facrul}
M_{n}(t_{1},\cdots,t_{n})=\langle\epsilon(t_{1})\cdots\epsilon(t_{n})\rangle=\langle\epsilon(t_{1})\epsilon(t_{2})\rangle\langle\epsilon(t_{3})\cdots\epsilon(t_{n})\rangle,
\end{equation}
for the order of the time instants $t_{1}>\cdot\cdot\cdot>t_{n}$ $(n\geq3)$.

To obtain the quantum master equation~(\ref{eq:masequ}) of the qubit system, we need to derive the exact expressions of the phase angle $\Phi(t)$ and decay factor $\Gamma(t)$ based on Eq.~(\ref{eq:phadec}).
However, it is hardly to calculate them for general environmental noise due to the fact that the number of order of the cumulants approaches infinity.
It is difficult to derive the exact quantum master equation (first-order differential equation) for reduced density matrix $\rho_{S}(t)$ and only a few physical models can be exactly solved.
For the general case, we need to truncate the environmental correlations effects to some finite order within some approximations.
The time-ordered factorization rule for the higher-order moments in Eq.~(\ref{eq:facrul}) results in
all the higher-order moments closely associated with the second-order moments, which makes it possible to derive
the closed higher-order differential equation for reduced density matrix $\rho_{S}(t)$ to contain all the infinite environmental correlations effects.
As an effective methodology, we derive the closed higher-order differential equation for the reduced density matrix $\rho_{S}(t)$.
We can solve the closed higher-order differential equation to obtain the exact expression of the qubit reduced density matrix and to derive the exact solutions of the frequency shift $\phi(t)$ and decoherence rate $\gamma(t)$.

In terms of the statistical characteristics of the environmental noise $\epsilon(t)$ in Eqs.~(\ref{eq:avecor}) and~(\ref{eq:facrul}), the second-order moment of the environmental noise obeys the closed differential relation
\begin{equation}
\label{eq:clodifrel}
\frac{d^{2}}{dt^{2}}M_{2}(t,t')=-\kappa\frac{d}{dt}M_{2}(t,t')-2\eta\kappa M_{2}(t,t').
\end{equation}
Based on the closed differential relation for the moments in Eq.~(\ref{eq:clodifrel}) and the relation between cumulants and moments of the environmental noise
\begin{equation}
\label{eq:relcummom}
\begin{split}
C_{n}(t_{1},\cdots,t_{n})=&M_{n}(t_{1},\cdots,t_{n})-\sum_{k=1}^{n-1}{n-1\choose k-1}C_{k}(t_{1},\cdots,t_{k})\\
&\times M_{n-k}(t_{1},\cdots,t_{n-k}),
\end{split}
\end{equation}
with ${n\choose m}=n!/[(n-m)!m!]$, the dynamical evolution of the reduced density matrix of the single qubit system is governed by a closed third-order differential equation
\begin{equation}
\label{eq:clomasequ}
\begin{split}
\frac{d^{3}}{dt^{3}}\rho_{S}(t)=&-\kappa\Big[\frac{d^{2}}{dt^{2}}\rho_{S}(t)+2\eta\frac{d}{dt}\rho_{S}(t)\Big]\\
&-\frac{1}{4}\chi^{2}[\sigma_{z},[\sigma_{z},\frac{d}{dt}\rho_{S}(t)+\kappa\rho_{S}(t)]],
\end{split}
\end{equation}
where the initial conditions satisfy
\begin{equation}
\label{eq:inicons}
\begin{split}
  \frac{d}{dt}\rho_{S}(0)&=-\frac{i}{2}a_{0}\chi[\sigma_{z},\rho_{S}(0)],\\
  \quad\frac{d^{2}}{dt^{2}}\rho_{S}(0)&=-\frac{1}{4}\chi^{2}[\sigma_{z},[\sigma_{z},\rho_{S}(0)]].
\end{split}
\end{equation}
We have derived the closed third-order differential equation for the single qubit reduced density matrix in the presence of a nonequilibrium environment exhibiting non-Markovian and nonstationary RTN properties.
We stress that the reduced density matrix in the closed differential equation~(\ref{eq:clomasequ}) contains all the environmental correlations and it is exact without any approximation.
The exact solution of the reduced density matrix in Eq.~(\ref{eq:clomasequ}) can be analytically expressed as
\begin{equation}
\label{eq:sinreddenmat1}
  \rho_{S}(t)=\mathscr{L}^{-1}[\tilde{\rho}_{S}(s)], \tilde{\rho}_{S}(s)=\left
   (\begin{array}{cc}
    \rho_{11}(0)\frac{1}{s}&\rho_{10}(0)\tilde{D}^{*}(s)\\
    \rho_{01}(0)\tilde{D}(s)&\rho_{00}(0)\frac{1}{s}
   \end{array}
   \right),
\end{equation}
where $\tilde{D}(s)$ is Laplace transform of the decoherence function $D(t)$ given by
\begin{equation}
\label{eq:laplaceD}
\begin{split}
   \tilde{D}(s)&= \tilde{D}_{r}(s)+i\tilde{D}_{i}(s),\\
   \tilde{D}_{r}(s)&=\frac{s^{2}+\kappa s+2\eta\kappa}
   {s^{3}+\kappa s^{2}+(2\eta\kappa+\chi^{2})s+\kappa\chi^{2}},\\
   \tilde{D}_{i}(s)&=\frac{a_{0}\chi(s+\kappa)}
   {s^{3}+\kappa s^{2}+(2\eta\kappa+\chi^{2})s+\kappa\chi^{2}}.
\end{split}
\end{equation}
The exact expression of the decoherence function $D(t)$ can be obtained
based on the approach we derived in Refs.~\onlinecite{JChemPhys.149.094107,SciRep.10.88}.
We here focus on the general case that the denominator of $\tilde{D}(s)$ has three different roots $r_{n} (n=1,2,3)$
and the decoherence function $D(t)$ can be exactly expressed as
\begin{equation}
\label{eq:timeD}
\begin{split}
   D(t)&=D_{r}(t)+iD_{i}(t),\\
   D_{r}(t)&=\sum_{n=1}^{3}c_{r}^{n}e^{r_{n}t}, D_{i}(t)=\sum_{n=1}^{3}c_{i}^{n}e^{r_{n}t},
\end{split}
\end{equation}
where the coefficients satisfy $c_{n}^{r}=\lim\limits_{s\rightarrow r_{n}}\big[(s-r_{n})\tilde{D}_{r}(s)\big]$ and $c_{n}^{i}=\lim\limits_{s\rightarrow r_{n}}\big[(s-r_{n})\tilde{D}_{i}(s)\big]$.
Correspondingly, the frequency shift and decoherence rate can be respectively written as
\begin{equation}
\label{eq:freshidecrat}
\begin{split}
   \phi(t)&=-\frac{\sum\limits_{n=1}^{3}\sum\limits_{k=1}^{3}(c_{r}^{n}c_{i}^{k}-c_{i}^{n}c_{r}^{k})r_{k}e^{(r_{n}+r_{k})t}}
   {\Big(\sum\limits_{n=1}^{3}c_{r}^{n}e^{r_{n}t}\Big)^{2}+\Big(\sum\limits_{n=1}^{3}c_{i}^{n}e^{r_{n}t}\Big)^{2}},\\
   \gamma(t)&=-\frac{\sum\limits_{n=1}^{3}\sum\limits_{k=1}^{3}(c_{r}^{n}c_{r}^{k}+c_{i}^{n}c_{i}^{k})r_{k}e^{(r_{n}+r_{k})t}}
   {\Big(\sum\limits_{n=1}^{3}c_{r}^{n}e^{r_{n}t}\Big)^{2}+\Big(\sum\limits_{n=1}^{3}c_{i}^{n}e^{r_{n}t}\Big)^{2}}.
\end{split}
\end{equation}


\subsection{Disentanglement dynamics of a two qubit system}
In the following, we study the disentanglement dynamics of a two qubit system $T$ consisting of two noninteracting identical single qubits independently coupled to its local nonequilibrium environment.
With respect to the two qubit system Hamiltonian $H_{T}=(\hbar/2)\omega_{S}\sigma_{z}\otimes I_{2}+(\hbar/2)\omega_{S}I_{2}\otimes\sigma_{z}$ and the environment Hamiltonian $H_{E}$, the Hamiltonian of the total system $T+E$ in the interaction picture can be expressed as
\begin{equation}
\label{eq:twointHam}
  H_{TE}(t)=\frac{\hbar}{2}\epsilon(t)\sigma_{z}\otimes I_{2}+\frac{\hbar}{2}\epsilon(t)I_{2}\otimes\sigma_{z}.
\end{equation}

We construct the reduced density matrix of the two qubit system in the standard product basis $\mathcal{B}_{T}=\{|1\rangle=|11\rangle, |2\rangle=|10\rangle, |3\rangle=|01\rangle, |4\rangle=|00\rangle\}$.
Based on the two qubit basis and by taking a partial trace over the environmental degrees of freedom,
the reduced density matrix of the two qubit system can be expressed as
\begin{equation}
\label{eq:tworeddenmat}
\begin{split}
  \rho_{T}(t)&=\left\langle U_{TE}\bm(t;\epsilon(t)\bm)\rho_{T}(0)U_{TE}^{\dag}\bm(t;\epsilon(t)\bm)\right\rangle\\
  &=\sum_{\mu=1}^{4}K_{T\mu}(t)\rho_{T}(0)K_{T\mu}^{\dag}(t),
\end{split}
\end{equation}
where $U_{TE}\bm(t;\epsilon(t)\bm)=\exp\big[-(i/\hbar)\int_{0}^{t}dt'H_{TE}(t')\big]$
denotes the unitary time evolution operator in terms of the Hamiltonian $H_{TE}(t)$ in Eq.~(\ref{eq:twointHam})
and the two qubit Kraus operators $K_{T\mu}(t)=K_{S\nu}(t)\otimes K_{S\upsilon}(t) (\nu,\upsilon=1,2)$ are the tensor products of the single qubit Kraus operators
\begin{equation}
\label{eq:twoqubKra}
\begin{split}
  K_{T1}(t)&=\left(\begin{array}{cc}
    1&0\\
    0&D(t)
   \end{array}
   \right)\otimes\left(\begin{array}{cc}
    1&0\\
    0&D(t)
   \end{array}
   \right),\\
   K_{T2}(t)&=\left(\begin{array}{cc}
    1&0\\
    0&D(t)
   \end{array}
   \right)\otimes\left(\begin{array}{cc}
    1&0\\
    0&\sqrt{1-|D(t)|^{2}}
   \end{array}
   \right),\\
   K_{T3}(t)&=\left(\begin{array}{cc}
    0&0\\
    0&\sqrt{1-|D(t)|^{2}}
   \end{array}
   \right)\otimes\left(\begin{array}{cc}
    1&0\\
    0&D(t)
   \end{array}
   \right),\\
   K_{T4}(t)&=\left(\begin{array}{cc}
    0&0\\
    0&\sqrt{1-|D(t)|^{2}}
   \end{array}
   \right)\otimes\left(\begin{array}{cc}
    0&0\\
    0&\sqrt{1-|D(t)|^{2}}
   \end{array}
   \right).
\end{split}
\end{equation}
According to the two qubit Kraus operators expression for the reduced density matrix in Eq.~(\ref{eq:tworeddenmat}), the diagonal elements are time independent
\begin{equation}
\label{eq:twodigele}
\begin{split}
  \rho_{11}(t)&=\rho_{11}(0),\\
  \rho_{22}(t)&=\rho_{22}(0),\\
  \rho_{33}(t)&=\rho_{33}(0),\\
  \rho_{44}(t)&=1-[\rho_{11}(0)+\rho_{22}(0)+\rho_{33}(0)],
\end{split}
\end{equation}
and time-dependent off diagonal elements can be written as
\begin{equation}
\label{eq:twooffdigele}
\begin{split}
  \rho_{21}(t)&=\rho^{*}_{12}(t)=\rho_{21}(0)D(t),\\
  \rho_{31}(t)&=\rho^{*}_{13}(t)=\rho_{31}(0)D(t),\\
  \rho_{32}(t)&=\rho^{*}_{23}(t)=\rho_{32}(0)|D(t)|^{2},\\
  \rho_{41}(t)&=\rho^{*}_{14}(t)=\rho_{41}(0)D^{2}(t),\\
  \rho_{42}(t)&=\rho^{*}_{24}(t)=\rho_{42}(0)D(t),\\
  \rho_{43}(t)&=\rho^{*}_{34}(t)=\rho_{43}(0)D(t).
\end{split}
\end{equation}

By taking the optimization over all pairs of initial states, the non-Markovianity quantifying the flow of information exchange between the two qubit system and environment can be expressed as~\cite{PhysRevLett.103.210401,PhysRevA81.062115}
\begin{equation}
\label{eq:twononMar}
\begin{split}
 \mathcal{N}_{T}&=\max\limits_{\rho_{T}^{1,2}(0)}\int_{\frac{d\mathcal{D}}{dt}>0}\frac{d}{dt}
 \mathcal{D}(\rho_{T}^{1}(t),\rho_{T}^{2}(t))dt\\
 &=-2\int_{\gamma(t)<0}\gamma(t)|D(t)|^{2}dt,
\end{split}
\end{equation}
where the optimal pair of initial states can be chosen as the maximally entangled states super-decoherent Bell states $|\psi_{T}^{\pm}(0)\rangle=(|00\rangle\pm|11\rangle)/\sqrt{2}$ or sub-decoherent Bell states $|\varphi_{T}^{\pm}(0)\rangle=(|01\rangle\pm|10\rangle)/\sqrt{2}$~\cite{PhysRevA87.052109,PhysRevA90.052103}.
It is obvious that similar to the single qubit case, the two qubit dynamics also displays non-Markovian behavior if the decoherence rate $\gamma(t)$ takes negative values sometimes which depends on both the environmental coupling $\chi$ and memory decay rate $\kappa$.
Because of the entanglement between the two identical single qubits for the optimal pair of maximally entangled states, the non-Markovianity of the two qubit system is less than twice that of the single qubit system, namely, $\mathcal{N}_{T}<2\mathcal{N}_{S}$.

Due to the environmental effects on its evolution, the two qubit system undergoes dynamical disentanglement.
For a two qubit system, all the entanglement measures are compatible.
In order to study the nonstationary and non-Markovian effects of the nonequilibrium environment on the disentanglement dynamics
of the two qubit system, we here use the concurrence to quantify the entanglement defined as~\cite{PhysRevLett.80.2245,PhysRevLett.93.140404}
\begin{equation}
\label{eq:concur}
\mathscr{C}(t)=\max\left\{0, \sqrt{\lambda_{1}(t)}-\sqrt{\lambda_{2}(t)}-\sqrt{\lambda_{3}(t)}-\sqrt{\lambda_{4}(t)}\right\},
\end{equation}
where $\lambda_{i}(t)$ are the eigenvalues of the matrix $\zeta(t)=\rho_{T}(t)(\sigma_{y}\otimes\sigma_{y})\rho_{T}^{*}(t)(\sigma_{y}\otimes\sigma_{y})$
arranged in decreasing order with $\rho_{T}^{*}(t)$ denoting the complex conjugation of the
two qubit reduced density matrix $\rho_{T}(t)$ in the two qubit basis $\mathcal{B}_{T}$.
The concurrence $\mathscr{C}(t)$ varies from the maximum 1 for a maximally entangled state to the minimum 0 for a completely disentangled state.
For pure quantum state, the entanglement corresponds to nonlocal correlations whereas it is not the general case for mixed states  due to the fact that the environmental noise gives rise to the decay of nonlocal correlations.
The nonlocality can be identified by the violation of the Bell inequalities in the presence of entanglement ($\mathscr{C}(t)>0$).
The Clauser-Horne-Shimony-Holt (CHSH) form of Bell function has been widely used to determine whether there are nonlocal correlations of the entangled system.
The maximum Bell function $\mathscr{B}(t)$ for an entangled two qubit system can be, based on the Horodecki criterion, expressed as~\cite{PhysLettA200.340}
\begin{equation}
\label{eq:Belfun}
\mathscr{B}(t)=2\sqrt{\max_{j>k}[\mu_{j}(t)+\mu_{k}(t)]},
\end{equation}
where the subscripts $j,k=1,2,3$ and $\mu_{j}(t)$ and $\mu_{k}(t)$ are functions in terms of the elements of the two qubit reduced density matrix.
If the Bell function $\mathscr{B}(t)$ is larger than the classical threshold $\mathscr{B}_{\mathrm{th}}=2$, the quantum correlations of the entangled two qubit system cannot be reproduced by any classical local model.

It is known that the Bell states and Werner mixed states of a two qubit system play an essential role in quantum computation and quantum information~\cite{Nielsenbook}.
The two qubit reduced density matrix expressed in Eq.~(\ref{eq:tworeddenmat}) for initial composite Bell states and extended Werner states has an $X$ structure both initially and during the dynamical evolution.
The concurrence $\mathscr{C}(t)$ for an initial $X$ structure reduced density matrix of a two qubit system can be computed in a particular form as~\cite{QuantumInfComput.7.459}
\begin{equation}
\label{eq:concurX}
\mathscr{C}_{X}(t)=\max\left\{0,\mathscr{C}_{1}(t),\mathscr{C}_{2}(t)\right\},
\end{equation}
where
\begin{equation}
\label{eq:concurcomX}
\begin{split}
\mathscr{C}_{1}(t)&=2\Big[|\rho_{23}(t)|-\sqrt{\rho_{11}(t)\rho_{44}(t)}\Big],\\
\mathscr{C}_{2}(t)&=2\Big[|\rho_{14}(t)|-\sqrt{\rho_{22}(t)\rho_{33}(t)}\Big].
\end{split}
\end{equation}
The time dependent maximum CHSH-Bell function $\mathscr{B}(t)$ for an X structure two qubit density matrix can be expressed  analytically as~\cite{PhysRevA72.042321,PhysRevA81.052116}
\begin{equation}
\label{eq:BelfunX}
\mathscr{B}_{X}(t)=\max\left\{\mathscr{B}_{1}(t),\mathscr{B}_{2}(t)\right\},
\end{equation}
where $\mathscr{B}_{1}(t)=2\sqrt{\mu_{1}(t)+\mu_{2}(t)}$ and $\mathscr{B}_{2}(t)=2\sqrt{\mu_{1}(t)+\mu_{3}(t)}$ with
\begin{equation}
\label{eq:Belcoes}
\begin{split}
\mu_{1}(t)&=4\left[|\rho_{14}(t)|+|\rho_{23}(t)|\right]^{2},\\
\mu_{2}(t)&=\left[\rho_{11}(t)+\rho_{44}(t)-\rho_{22}(t)-\rho_{33}(t)\right]^{2},\\
\mu_{3}(t)&=4\left[|\rho_{14}(t)|-|\rho_{23}(t)|\right]^{2}.
\end{split}
\end{equation}

We first focus on the initial states of the system in the composite Bell states of the form~\cite{PhysRevLett.104.200401}
\begin{equation}
\label{eq:comBelsta}
  \rho_{T}(0)=\frac{1+c}{2}|\psi_{T}^{\pm}(0)\rangle\langle\psi_{T}^{\pm}(0)|
  +\frac{1-c}{2}|\varphi_{T}^{\pm}(0)\rangle\langle\varphi_{T}^{\pm}(0)|,
\end{equation}
where the initial state parameter $c$ is real and satisfies $-1\leq c\leq1$.
It has, by studying the quantum mutual information, quantum discord and classical correlations of the dynamics, shown that for the initial states in Eq.~(\ref{eq:comBelsta}) there is a sudden transition from classical to quantum decoherence for the two qubit system coupled to a nonequilibrium environment exhibiting with generalized RTN property, and the nonequilibrium feature of the environment can delay the critical time of the transition of decoherence from classical to quantum~\cite{NewJPhys.22.033039}.
The concurrence at time $t$ in Eq.~(\ref{eq:concur}) for the two qubit system prepared in the initial states of Eq.~(\ref{eq:comBelsta}) can be reduced to
\begin{equation}
\label{eq:concurCBS}
\mathscr{C}(t)=\max\left\{0,(1-c)|D(t)|^2-(1+c),(1+c)|D(t)|^2-(1-c)\right\}.
\end{equation}
The initial concurrence of the two qubit system prepared in the composite Bell states in Eq.~(\ref{eq:comBelsta}) can be
expressed as $\mathscr{C}(0)=2|c|$ since the initial value of the decoherence function satisfies $D(0)=1$.
Therefore, the entanglement of the two qubit system exists except for the special case $c=0$.
For the case $-1\leq c<0$, the concurrence at time $t$ exists if the threshold value of the decoherence function satisfies $|D(t)|>|D_{\mathrm{th}}|=\sqrt{(1+c)/(1-c)}$ whereas if it exists at time $t$ for the case $0<c\leq1$, the threshold value of the decoherence function satisfies $|D(t)|>|D_{\mathrm{th}}|=\sqrt{(1-c)/(1+c)}$.
The time dependent maximum CHSH-Bell function $\mathscr{B}(t)$ for the initial states of the two qubit system of Eq.~(\ref{eq:comBelsta}) can be reduced to
\begin{equation}
\label{eq:BelfunCBS}
\mathscr{B}(t)=4\sqrt{|D(t)|^{4}+c^{2}}.
\end{equation}
The presence of entanglement $\mathscr{C}(t)>0$ is a necessary condition to achieve nonlocality.
The close relation between $\mathscr{B}(t)$ and $\mathscr{C}(t)$
for the two qubit system prepared in the initial composite Bell states of Eq.~(\ref{eq:comBelsta}) can be expressed as
\begin{equation}
\label{eq:BCrelCBS}
\mathscr{B}(t)=
\left\{\begin{array}{cc}
\frac{4}{1-c}\sqrt{\big[\mathscr{C}(t)+(1+c)\big]^{2}+c^{2}(1-c)^{2}},&-1\leq c<0,\\
\frac{4}{1+c}\sqrt{\big[\mathscr{C}(t)+(1-c)\big]^{2}+c^{2}(1+c)^{2}},&0<c\leq1.
\end{array} \right.
\end{equation}
For the case $-1\leq c<0$, the classical threshold $\mathscr{C}_{\mathrm{th}}$ which corresponds to the Bell function $\mathscr{B}(t)\geq\mathscr{B}_{\mathrm{th}}=2$ only exists for $-1/2\leq c<0$ and can be expressed as $\mathscr{C}_{\mathrm{th}}=(1-c)\sqrt{1/4-c^{2}}-(1+c)$ whereas for $-1\leq c<-1/2$ the maximum CHSH-Bell function $\mathscr{B}(t)$ is always larger than the threshold $\mathscr{B}_{\mathrm{th}}=2$.
Similarly, for the case $0<c\leq1$, the threshold $\mathscr{C}_{\mathrm{th}}$ for the Bell function larger than the threshold $\mathscr{B}_{\mathrm{th}}=2$ exists for $0<c\leq1/2$ and can be expressed as $\mathscr{C}_{\mathrm{th}}=(1+c)\sqrt{1/4-c^{2}}-(1-c)$ while the maximum CHSH-Bell function $\mathscr{B}(t)$ is always larger than the threshold $\mathscr{B}_{\mathrm{th}}=2$ for $1/2<c\leq1$.
Therefore, the two qubit system always displays quantum nonlocality at time $t$ for the case $1/2<|c|\leq1$
whereas for the case $0<|c|\leq1/2$ the threshold value of the decoherence function should satisfies $|D(t)|>|D_{\mathrm{th}}|=\sqrt[4]{1/2-c^{2}}$ to ensure that the concurrence $\mathscr{C}(t)$ is larger than the classical threshold $\mathscr{C}_{\mathrm{th}}$.

We now focus on the case that the two qubit system is prepared in the extended Werner states initially as~\cite{PhysRevA77.032342,PhysRevB90.054304}
\begin{equation}
\label{eq:extWersta}
\begin{split}
\rho_{T}^{\psi}(0)&=r|\psi_{T}(0)\rangle\langle\psi_{T}(0)|+\frac{1-r}{4}I_{4},\\
\rho_{T}^{\varphi}(0)&=r|\varphi_{T}(0)\rangle\langle\varphi_{T}(0)|+\frac{1-r}{4}I_{4},
\end{split}
\end{equation}
where $0\leq r\leq1$ denotes the purity parameter of the initial states, $I_{4}$ is the $4\times4$ identity matrix and $|\psi_{T}(0)\rangle=\alpha|00\rangle+\beta|11\rangle$ and $|\varphi_{T}(0)\rangle=\alpha|01\rangle+\beta|10\rangle$ represent the initial entangled states with the complex coefficients $\alpha$, $\beta$ and $|\alpha|^{2}+|\beta|^{2}=1$.
For the case $\alpha=\pm\beta=1/\sqrt{2}$, the extended Werner states corresponds to a subclass of
Bell-diagonal states, namely, the Werner states~\cite{PhysRevLett.89.170401,RevModPhys.81.865}.
The concurrence for the two qubit system prepared in the extended Werner states initially of Eq.~(\ref{eq:extWersta}) can be reduced to
\begin{equation}
\label{eq:concurEWS}
\mathscr{C}^{\psi}(t)=\mathscr{C}^{\varphi}(t)=\max\left\{0,2|\alpha\beta||D(t)|^2-\frac{1}{2}(1-r)\right\}.
\end{equation}
The entanglement of the two qubit system exists if the initial value of concurrence $\mathscr{C}(0)$ in the extended Werner states is larger than zero, correspondingly
\begin{equation}
\label{eq:cripur}
r>\frac{1}{4|\alpha\beta|+1},
\end{equation}
except for the special cases $\alpha=0$ and $\alpha=1$.
The concurrence exists at time $t$ if the threshold value of the decoherence function satisfies $|D(t)|>|D_{\mathrm{th}}|=\sqrt{(1-r)/|\alpha\beta|}/2$.
The time dependent maximum CHSH-Bell function $\mathscr{B}(t)$ for the initial extended Werner states of Eq.~(\ref{eq:extWersta}) can be reduced to
\begin{equation}
\label{eq:BelfunEWS}
\mathscr{B}(t)=2r\sqrt{1+4|\alpha\beta|^{2}|D(t)|^{4}}.
\end{equation}
In the presence of entanglement $\mathscr{C}(t)>0$, for the two qubit system prepared initially in the extended Werner states of Eq.~(\ref{eq:extWersta}), the close relation between $\mathscr{B}(t)$ and $\mathscr{C}(t)$ can be expressed as
\begin{equation}
\label{eq:BCrelEWS}
\mathscr{B}(t)=2\sqrt{r^{2}+\Big[\mathscr{C}(t)+\frac{1}{2}(1-r)\Big]^{2}}.
\end{equation}
The classical threshold $\mathscr{C}_{\mathrm{th}}$ corresponding to the Bell function $\mathscr{B}(t)\geq\mathscr{B}_{\mathrm{th}}=2$ can be expressed as $\mathscr{C}_{\mathrm{th}}=\sqrt{1-r^{2}}-(1-r)/2$ which depends only on the purity parameter $r$ of the initial states of Eq.~(\ref{eq:extWersta}), and it is a decreasing function of the purity parameter $r$ and for the presence of entanglement $1/(4|\alpha\beta|+1)<r\leq1$ ($\alpha\neq0$ and $\alpha\neq1$), it satisfies $0\leq\mathscr{C}_{\mathrm{th}}<2\big(\sqrt{4|\alpha\beta|^{2}+2|\alpha\beta|}-|\alpha\beta|\big)/(4|\alpha\beta|+1)$.
Thus, the two qubit system displays quantum nonlocality at time $t$ provided that the concurrence $\mathscr{C}(t)$ is larger than the classical threshold $\mathscr{C}_{\mathrm{th}}$ corresponding to that the threshold value of the decoherence function satisfies $|D(t)|>|D_{\mathrm{th}}|=\sqrt[4]{1-r^{2}}/\sqrt{2|\alpha\beta|}$.

\section{Results and discussion}
\label{sec:result}

In this section we show the numerical results of the non-Markovian dynamics of a single qubit system and a two noninteracting qubit system coupled to nonequilibrium environments with nonstationary and non-Markovian RTN statistical properties, respectively.
In the presence of RTN only exhibiting Markovian feature, there are two important dynamics regimes identified: the weak coupling regime $\chi/\eta<1$ and the strong coupling regime $\chi/\eta>1$~\cite{PhysRevB75.054515,PhysRevB77.174509}, and the quantum dynamics displays Markovian and non-Markovian behavior in the weak coupling regime and strong coupling regime, respectively.
However, the boundary of the quantum dynamics regimes depends both on the environmental coupling $\chi$ and the environmental non-Markovian feature $\kappa$ in the presence of RTN exhibiting non-Markovian feature, and in the weak coupling regime, the quantum dynamics can also display non-Markovian behavior~\cite{PhysRevA94.042110,JChemPhys.149.094107,SciRep.10.88}.

\subsection{Decoherence of a single qubit system in nonequilibrium environments}
In this subsection, we show the numerical results of the non-Markovian decoherence dynamics of a single qubit system coupled to nonequilibrium environments. We mainly focus on  how the environmental nonstationary and non-Markovian features influence the non-Markovianity $\mathcal{N}$ and the decoherence function. The comparisons with the environmental stationary and memoryless cases are also discussed.

\begin{figure}[ht]
\centering
    \epsfig{file=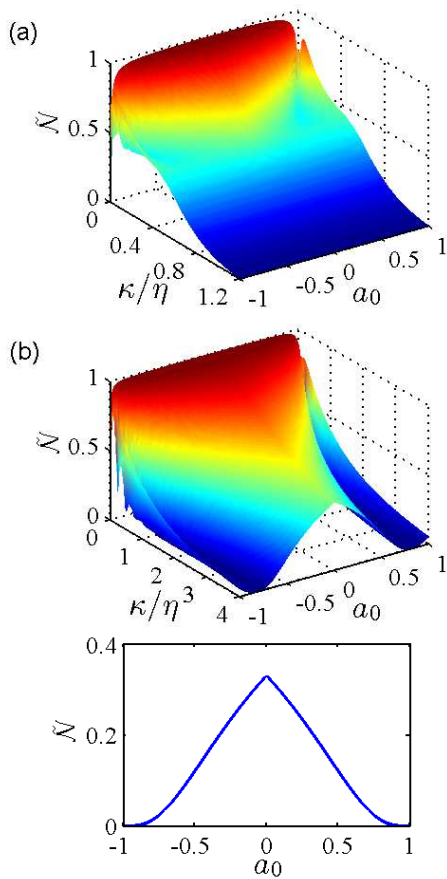,width=2.5in}
    \caption{(Color online) Scaled non-Markovianity $\tilde{\mathcal{N}}=\mathcal{N}/(\mathcal{N}+1))$ of a single qubit system in nonequilibrium environments as a function of the environmental memory decay rate $\kappa$ and the nonstationary parameter $a_{0}$ in (a) weak coupling regime with $\chi/\eta=0.8$ and (b) strong coupling regime with $\chi/\eta=3$.
    The bottom panel of (b) stands for the memoryless case $\kappa\rightarrow+\infty$.}
    \label{fig:nMsq}
\end{figure}

Figure~\ref{fig:nMsq} shows the non-Markovianity $\mathcal{N}$ of a single qubit system coupled to nonequilibrium environments as a function of the environmental memory decay rate $\kappa$ and the nonstationary parameter $a_{0}$.
In both weak and strong coupling regimes, for a given value of the environmental memory decay rate $\kappa$, the non-Markovianity $\mathcal{N}$ displays symmetrical behavior for positive and negative values of the environmental nonstationary parameter $a_{0}$. For small value of the environmental memory decay rate $\kappa$ (strong environmental non-Markovian feature), the non-Markovianity $\mathcal{N}$ shows nonmonotonical decay whereas it shows monotonical decay for large value of the environmental memory decay rate $\kappa$ as the environmental nonstationary parameter $|a_{0}|$ increases.
This reflects that the environmental nonstationary feature can reduce the non-Markovianity $\mathcal{N}$ in weak environmental non-Markovian feature while it contributes nonmonotonically to the non-Markovian behavior of the decoherence dynamics in strong environmental non-Markovian feature.
For a given value of the environmental nonstationary parameter $a_{0}$, the non-Markovianity $\mathcal{N}$ increases as  the environmental memory decay rate $\kappa$ decreases in both weak and strong coupling regimes. This indicates that the environmental non-Markovian feature can enhance the non-Markovian behavior of the decoherence dynamics.
In the weak coupling regime as displayed in Fig.~\ref{fig:nMsq} (a), the non-Markovianity $\mathcal{N}$ decreases to zero as the environmental memory decay rate $\kappa$ increases whereas it does not decrease to zero in the strong coupling regime as shown in Fig.~\ref{fig:nMsq} (b). This reflects that the boundary of the decoherence dynamics regimes depends closely on the environmental coupling $\chi$ and the environmental non-Markovian feature $\kappa$.

\begin{figure}[ht]
\centering
    \epsfig{file=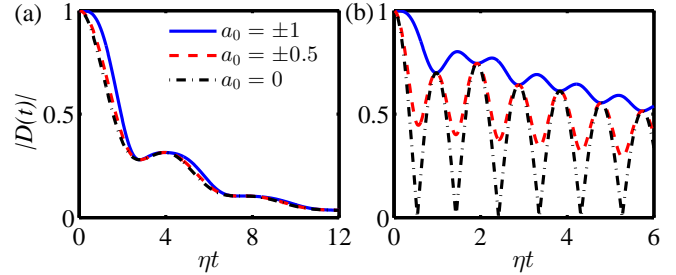,width=3.5in}
    \caption{(Color online) Time evolution of the decoherence function $|D(t)|$ for different values of the environmental nonstationary parameter $a_{0}$ in (a) weak coupling regime with $\chi/\eta=0.8$ and (b) strong coupling regime with $\chi/\eta=3$.
    The environmental memory decay rate is given by $\kappa/\eta=1$.
    The right panels of (a) and (b) are for the stationary case $a_{0}=0$.}
    \label{fig:dfsqa}
\end{figure}

Figure~\ref{fig:dfsqa} displays the time evolution of the decoherence function $|D(t)|$ for different values of the environmental nonstationary parameter $a_{0}$.
Obviously, the decoherence function $|D(t)|$ shows nonmonotonical decays with nonzero coherence revivals
for $a_{0}\neq0$ in both the weak and strong coupling regimes whereas for the stationary case $a_{0}=0$, it decays with nonzero coherence revivals and with zero coherence revivals in the weak and strong coupling regimes, respectively.
As the environmental nonstationary parameter $a_{0}$ departs from zero, the decoherence function $|D(t)|$ decays slowly in both the weak and strong coupling regimes. This implies that the environmental nonstationary feature can suppress the decoherence dynamics.
In the strong coupling regime as shown in Fig.~\ref{fig:dfsqa} (b), the nonmonotonical behavior in the decoherence function decreases as the environmental nonstationary parameter $a_{0}$ departs from zero.
In contrast, there is no obvious change in the weak coupling regime as shown in Fig.~\ref{fig:dfsqa} (a).
This reflects that the environmental nonstationary feature can reduce the coherence revivals in the strong coupling regime whereas it has no obvious influence on the coherence revivals in the weak coupling regime.
In addition, the environmental nonstationary feature can not make the dynamics regime of the qubit system undergo transition due to the fact that the boundary of the quantum dynamics regimes does not depend on the environmental nonstationary feature.

\begin{figure}[ht]
\centering
    \epsfig{file=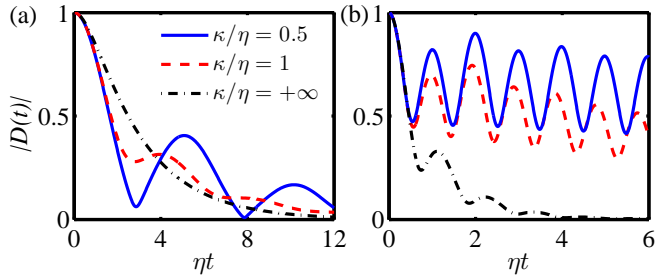,width=3.5in}
    \caption{(Color online) Time evolution of the decoherence function $|D(t)|$ for different values of the environmental memory decay rate $\kappa$ in (a) weak coupling regime with $\chi/\eta=0.8$ and (b) strong coupling regime with $\chi/\eta=3$.
    The environmental nonstationary parameter is given by $|a_{0}|=0.5$.
    The right panels of (a) and (b) stand for the Markovian case $\kappa=+\infty$.}
    \label{fig:dfsqkappa}
\end{figure}

Figure~\ref{fig:dfsqkappa} shows the time evolution of the decoherence function $|D(t)|$ for different values of the environmental memory decay rate $\kappa$.
In both the weak and strong coupling regimes, the coherence revivals in the decoherence function $|D(t)|$ enhance as the environmental memory decay rate $\kappa$ decreases.
This reflects that the environmental non-Markovian feature can enhance the non-Markovian behavior in the quantum dynamics.
Different from the strong coupling regime, the coherence revivals undergoes a transition from that with nonzeros to that with zeros in weak coupling regime as the environmental memory decay rate $\kappa$ decreases.
However, due to the difference in the coupling strength, the period of coherence revivals in the weak coupling regime is larger than that in the strong coupling regime.
For the environmental memoryless case $\kappa\rightarrow\infty$,  in the weak coupling regime as shown in Fig.~\ref{fig:dfsqkappa} (a), the decoherence function $|D(t)|$ decays monotonically with no coherence revivals whereas it decays nonmonotonically with nonzero coherence revivals in the strong coupling regime as shown in Fig.~\ref{fig:dfsqkappa} (b).
This indicates that the quantum dynamics displays a transition from Markovian to non-Markovian in the weak coupling regime as the environmental non-Markovian feature increases and the threshold value of the environmental memory decay rate $\kappa$ for the transition boundary depends on the value of the environmental coupling $\chi$.

\subsection{Disentanglement of a two qubit system in nonequilibrium environments}
In this subsection, we show the numerical results of the non-Markovian disentanglement dynamics of a two qubit system consisting of two noninteracting identical single qubits independently coupled to its local nonequilibrium environment. We mainly focus on how the environmental nonstationary and non-Markovian features influence the non-Markovianity $\mathcal{N}$, the entanglement quantified by the concurrence and the nonlocality characterized by the Bell function. The comparisons with the environmental stationary and memoryless cases are also discussed.

\begin{figure}[ht]
\centering
    \epsfig{file=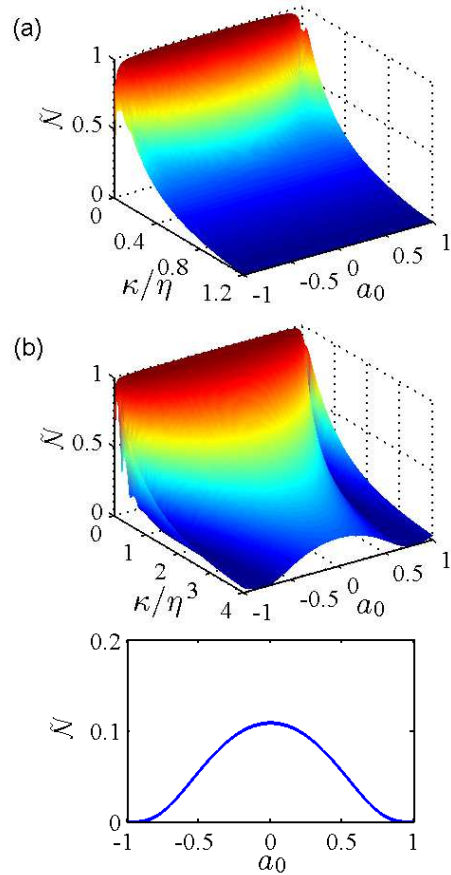,width=2.5in}
    \caption{(Color online) Scaled non-Markovianity $\tilde{\mathcal{N}}=\mathcal{N}/(\mathcal{N}+1))$ of a two qubit system in nonequilibrium environments as a function of the environmental memory decay rate $\kappa$ and the nonstationary parameter $a_{0}$ in (a) weak coupling regime with $\chi/\eta=0.8$ and (b) strong coupling regime with $\chi/\eta=3$.
    The bottom panel of (b) is for the memoryless case $\kappa\rightarrow+\infty$.}
    \label{fig:nMtq}
\end{figure}

Figure~\ref{fig:nMtq} shows the non-Markovianity $\mathcal{N}$ of a two qubit system interacting with nonequilibrium environments as a function of the environmental memory decay rate $\kappa$ and the nonstationary parameter $a_{0}$.
Similar to the case of a single qubit system coupled to nonequilibrium environments, for a given value of the environmental memory decay rate $\kappa$, the non-Markovianity $\mathcal{N}$ shows symmetrical behavior for positive and negative environmental nonstationary parameter $a_{0}$ in both weak and strong coupling regimes.
As the environmental nonstationary parameters $|a_{0}|$ deviates from zero, the non-Markovianity $\mathcal{N}$ displays nonmonotonical and monotonical decays for small and large $\kappa$, respectively.
In both weak and strong coupling regimes, for a given value of the environmental nonstationary parameter $a_{0}$, the non-Markovianity $\mathcal{N}$ increases with the decrease of the environmental memory decay rate $\kappa$.
The non-Markovianity $\mathcal{N}$ decreases to zero as the environmental memory decay rate $\kappa$ increases in the weak coupling regime as shown in Fig.~\ref{fig:nMtq} (a) whereas it does not decrease to zero in the strong coupling regime as displayed in Fig.~\ref{fig:nMtq} (b).

\begin{figure}[ht]
\centering
    \epsfig{file=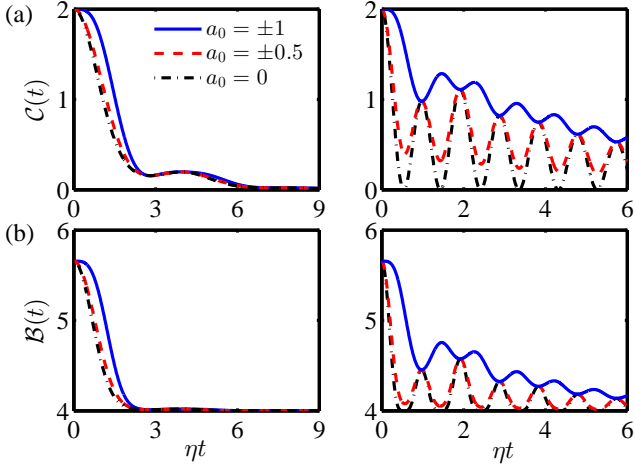,width=3.5in}
    \caption{(Color online) Time evolution of (a) the concurrence $\mathscr{C}(t)$ and (b) the Bell function $\mathscr{B}(t)$ for different values of the environmental nonstationary parameter $a_{0}$ for the two qubit system prepared initially in the composite Bell states with the initial state parameter $|c|=1$. Left panel: the weak coupling regime with $\chi/\eta=0.8$. Right panel: the strong coupling regime with $\chi/\eta=3$.
    The environmental memory decay rate is given by $\kappa/\eta=1$.}
    \label{fig:CBaBS}
\end{figure}

Figure~\ref{fig:CBaBS} displays the time evolution of the concurrence $\mathscr{C}(t)$ and the Bell function $\mathscr{B}(t)$ for different values of the environmental nonstationary parameter $a_{0}$ for the two qubit system prepared initially in the composite Bell states.
As shown in Fig.~\ref{fig:CBaBS} (a), the concurrence $\mathscr{C}(t)$ decays nonmonotonically with nonzero entanglement revivals in the weak coupling regime for both the nonstationary $a_{0}\neq0$ and stationary $a_{0}=0$ cases. In contrast, in the strong coupling regime the concurrence $\mathscr{C}(t)$ decays nonmonotonically with nonzero entanglement revivals and zero entanglement revivals for the nonstationary case $a_{0}\neq0$ and stationary case $a_{0}=0$, respectively.
In both the weak and strong coupling regimes, the concurrence $\mathscr{C}(t)$ increases as the environmental nonstationary parameter $a_{0}$ departs from zero.
This indicates that the environmental nonstationary feature can suppress the disentanglement dynamics.
In both the weak and strong coupling regimes as displayed in Fig.~\ref{fig:CBaBS} (b), the Bell function $\mathscr{B}(t)$ shows nonmonotonical decays with oscillations and it increases as the environmental nonstationary parameter $a_{0}$ departs from zero.
This reflects that the environmental nonstationary feature can enhance the quantum nonlocality.

\begin{figure}[ht]
\centering
    \epsfig{file=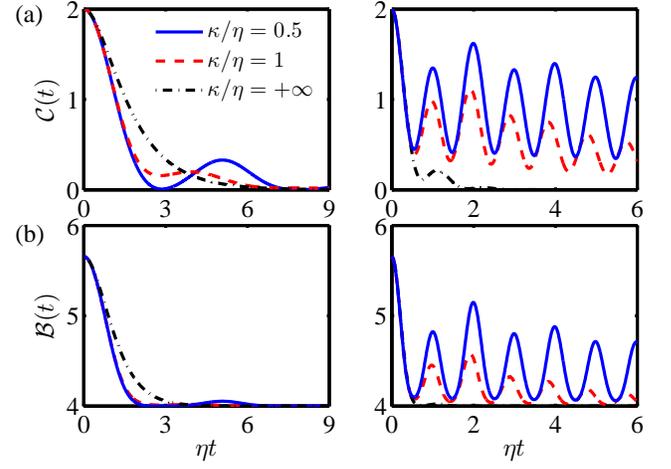,width=3.5in}
    \caption{(Color online) Time evolution of (a) the concurrence $\mathscr{C}(t)$ and (b) the Bell function $\mathscr{B}(t)$ for different values of the environmental memory decay rate $\kappa$ for the two qubit system prepared initially in the composite Bell states with the initial state parameter $|c|=1$. Left panel: the weak coupling regime with $\chi/\eta=0.8$. Right panel: the strong coupling regime with $\chi/\eta=3$.
    The environmental nonstationary parameter is given by $|a_{0}|=0.5$.}
    \label{fig:CBkappaBS}
\end{figure}

Figure~\ref{fig:CBkappaBS} displays the time evolution of the concurrence $\mathscr{C}(t)$ and the Bell function $\mathscr{B}(t)$ for different values of the environmental memory decay rate $\kappa$ for the two qubit system prepared initially in the composite Bell states.
As shown in Fig.~\ref{fig:CBkappaBS} (a), the concurrence $\mathscr{C}(t)$ decays nonmonotonically with nonzero entanglement revivals in the strong coupling regime whereas in the weak coupling regime, it undergoes a transition from nonmonotonical decay to monotonical decay as the environmental memory decay rate $\kappa$ increases.
The entanglement revivals in the concurrence $\mathscr{C}(t)$ become obvious as the environmental memory decay rate $\kappa$ decreases in both the weak and strong coupling regimes.
This indicates that the environmental non-Markovian feature can enhance the entanglement revivals and suppress the disentanglement dynamics.
As displayed in Fig.~\ref{fig:CBkappaBS} (b), in the strong coupling regime the Bell function $\mathscr{B}(t)$ decays nonmonotonically and it increases as the environmental memory decay rate $\kappa$ decreases.
This reflects that the environmental non-Markovian feature can enhance the quantum nonlocality in the strong coupling regime.
In contrast, the decay of the Bell function $\mathscr{B}(t)$ exhibits a transition from nonmonotonical decay to monotonical decay with the increase of the environmental memory decay rate $\kappa$ in the weak coupling regime.
In some time intervals, the Bell function $\mathscr{B}(t)$ decreases while it increases in some other time intervals as the environmental memory decay rate $\kappa$ decreases.

\begin{figure}[ht]
\centering
    \epsfig{file=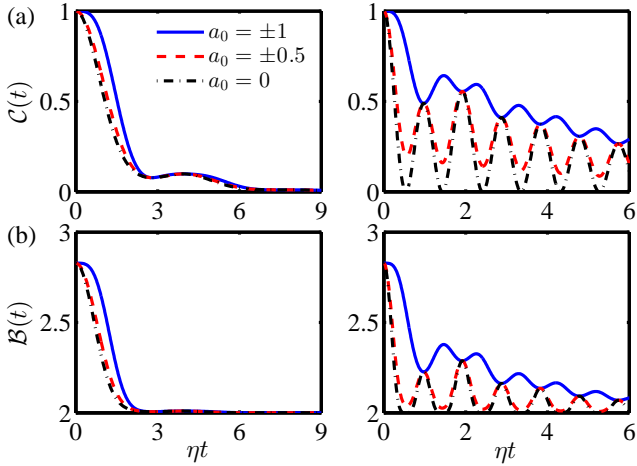,width=3.5in}
    \caption{(Color online) Time evolution of (a) the concurrence $\mathscr{C}(t)$ and (b) the Bell function $\mathscr{B}(t)$ for different values of the environmental nonstationary parameter $a_{0}$ for the two qubit system prepared initially in the extended Werner states with the initial purity parameter $r=1$ and the coefficients $|\alpha|=|\beta|=1/\sqrt{2}$. Left panel: the weak coupling regime with $\chi/\eta=0.8$. Right panel: the strong coupling regime with $\chi/\eta=3$.
    The environmental memory decay rate is given by $\kappa/\eta=1$.}
    \label{fig:CBaEWS}
\end{figure}

Figure~\ref{fig:CBaEWS} shows the time evolution of the concurrence $\mathscr{C}(t)$ and the Bell function $\mathscr{B}(t)$ for different values of the environmental nonstationary parameter $a_{0}$ for the two qubit system prepared initially in the extended Werner states.
Similar to the case that the two qubit system initially prepared in the composite Bell states, as displayed in Fig.~\ref{fig:CBaEWS} (a), the concurrence $\mathscr{C}(t)$ decays nonmonotonically with nonzero entanglement revivals in both the weak and strong coupling regimes for the nonstationary $a_{0}\neq0$.
However, for the stationary case $a_{0}=0$, the concurrence $\mathscr{C}(t)$ decays nonmonotonically with nonzero and zero entanglement revivals in the weak and strong coupling regimes, respectively.
In addition, the concurrence $\mathscr{C}(t)$ increases as the environmental nonstationary parameter $a_{0}$ departs from zero in both the weak and strong coupling regimes.
As shown in Fig.~\ref{fig:CBaEWS} (b), the Bell function $\mathscr{B}(t)$ shows nonmonotonical decays with oscillations and it increases as the environmental nonstationary parameter $a_{0}$ departs from zero in both the weak and strong coupling regimes.
These results indicate that the environmental nonstationary feature can suppress the disentanglement dynamics and enhance the quantum nonlocality.
Comparing Fig.~\ref{fig:CBaEWS} with Fig.~\ref{fig:CBaBS}, the trends of the influence of the environmental nonstationary feature on the disentanglement dynamics and quantum nonlocality do not depend on the initial states.

\begin{figure}[ht]
\centering
    \epsfig{file=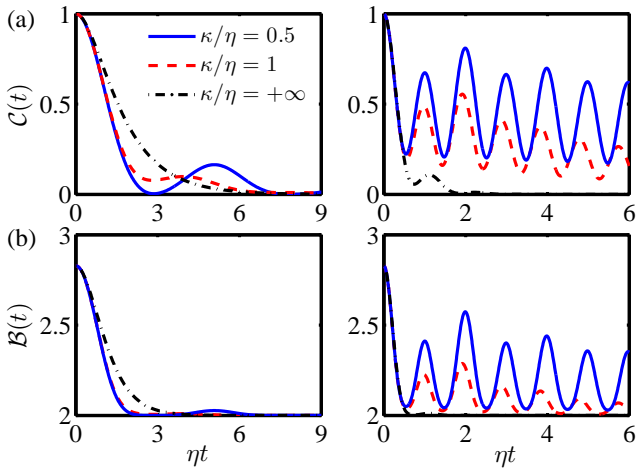,width=3.5in}
    \caption{(Color online) Time evolution of (a) the concurrence $\mathscr{C}(t)$ and (b) the Bell function $\mathscr{B}(t)$ for different values of the environmental memory decay rate $\kappa$ for the two qubit system prepared initially in the extended Werner states with the initial purity parameter $r=1$ and the coefficients $|\alpha|=|\beta|=1/\sqrt{2}$. Left panel: the weak coupling regime with $\chi/\eta=0.8$. Right panel: the strong coupling regime with $\chi/\eta=3$.
    The environmental nonstationary parameter is given by $|a_{0}|=0.5$.}
    \label{fig:CBkappaEWS}
\end{figure}

Figure~\ref{fig:CBkappaEWS} displays the time evolution of the concurrence $\mathscr{C}(t)$ and the Bell function $\mathscr{B}(t)$ for different values of the environmental memory decay rate $\kappa$ for the two qubit system prepared initially in the extended Werner states.
As displayed in Fig.~\ref{fig:CBkappaEWS} (a), similar to the case that the two qubit system initially prepared in the composite Bell states, the concurrence $\mathscr{C}(t)$ decays nonmonotonically with nonzero entanglement revivals in the strong coupling regime whereas it exhibits a transition from nonmonotonical decay to monotonical decay  in the weak coupling regime as the environmental memory decay rate $\kappa$ increases.
In both the weak and strong coupling regimes, the entanglement revivals in the concurrence $\mathscr{C}(t)$ enhances as the environmental memory decay rate $\kappa$ decreases.
As shown in Fig.~\ref{fig:CBkappaBS} (b), the Bell function $\mathscr{B}(t)$ decays nonmonotonically and it increases with the increase of the environmental memory decay rate $\kappa$ in the strong coupling regime.
In contrast, in the weak coupling regime the Bell function $\mathscr{B}(t)$ exhibits a transition from nonmonotonical decay to monotonical decay as the environmental memory decay rate $\kappa$ decreases and the Bell function $\mathscr{B}(t)$ decreases in some time intervals and increases in some other time intervals as the environmental memory decay rate $\kappa$ decreases.
Comparing Fig.~\ref{fig:CBkappaEWS} with Fig.~\ref{fig:CBkappaBS}, similar to the influence of the environmental nonstationary feature, the trends of the influence of the environmental non-Markovian feature on the disentanglement dynamics and quantum nonlocality do not depend on the initial states.

\section{Conclusions}
\label{sec:con}

We have theoretically studied the non-Markovian decoherence dynamics of a single qubit system and disentanglement dynamics of a two qubit system in the presence of nonequilibrium environments with nonstationary and non-Markovian statistical properties.
We have derived the closed third-order differential equation for the single qubit reduced density matrix with the nonequilibrium environments exhibiting nonstationary and non-Markovian RTN properties.
The reduced density matrix of the two qubit system can be expressed in terms of the Kraus representation by means of the tensor products of the single qubit Kraus operators.
We have derived the relation between the entanglement characterized by the concurrence and nonlocality quantified by by the Bell function of the two qubit system which are both closely associated with the decoherence function and
we have identified the threshold values of the decoherence function to ensure the existences of the concurrence and nonlocal quantum correlations for a given evolution time when the two qubit system is initially prepared in the composite Bell states and the extended Werner states, respectively.
The results show that the decoherence dynamics of a single qubit system and the disentanglement dynamics of a two qubit system can be suppressed by the environmental nonstationary feature.
The environmental nonstationary feature can reduce the coherence revivals in the single decoherence dynamics and entanglement revivals in the two qubit disentanglement dynamics.
Furthermore, the environmental non-Markovian feature can increase the coherence and entanglement revivals in the single and two qubit dynamics, respectively.
Moreover, the environmental nonstationary and non-Markovian features can enhance the nonlocality of the two qubit system.
Our results are helpful for further understanding the non-Markovian qubit dynamics in the presence of nonequilibrium environments.

\begin{acknowledgments}
  This work was supported by the National Natural Science Foundation of China (Grant No.11947033).
X.C. also acknowledges the support from the Doctoral Research Fund of Shandong Jianzhu University (Grant No. XNBS1852).
\end{acknowledgments}

\end{document}